\documentclass[12pt]{article}
\usepackage{graphicx}

\newcommand\pubnumber{Article 15 in eConf C1304143}
\newcommand\pubdate{\today}

\def\Title#1{\begin{center} {\Large #1 } \end{center}}
\def\Author#1{\begin{center}{ \sc #1} \end{center}}
\def\Address#1{\begin{center}{ \it #1} \end{center}}

\newcommand\pubblock{\rightline{\begin{tabular}{l} \pubnumber\\
         \pubdate  \end{tabular}}}
\newenvironment{Abstract}{\begin{quotation}  }{\end{quotation}}
\newenvironment{Presented}{\begin{quotation} \begin{center}
             PRESENTED AT\end{center}\bigskip
      \begin{center}\begin{large}}{\end{large}\end{center} \end{quotation}}

\begin{document}
\begin{titlepage}
\pubblock

\vfill
\Title{X - Ray Flares and Their Connection With Prompt Emission in GRBs}
\vfill
\Author{E. Sonbas$^{1,2}$, G. A. MacLachlan$^{3}$, A. Shenoy$^{3}$, K.S. Dhuga$^{3}$, \\
W. C. Parke$^{3}$}
\Address{{$^1$University of Adiyaman, Department of Physics, 02040 Adiyaman, Turkey} \\
{$^2$NASA Goddard Space Flight Center, Greenbelt, MD 20771, USA} \\
{$^3$Department of Physics, The George Washington University, Washington, DC 20052, USA} \\}
\vfill
\begin{Abstract}
We use a wavelet technique to investigate the time variations in the light curves from a sample of GRBs detected by Fermi and Swift. We focus 
primarily on the behavior of the flaring region of Swift-XRT light curves in order to explore connections between variability time scales and 
pulse parameters (such as rise and decay times, widths, strengths, and separation distributions) and spectral lags. Tight correlations between some 
of these temporal features suggest a common origin for the production of X-ray flares and the prompt emission.  
\end{Abstract}
\vfill
\begin{Presented}
GRB 2013 \\
the Seventh Huntsville Gamma-Ray Burst Symposium \\
Nashville, Tennessee, 14--18 April 2013
\end{Presented}
\vfill
\end{titlepage}
\def\thefootnote{\fnsymbol{footnote}}
\setcounter{footnote}{0}

\section{Introduction}

In addition to the prompt emission in Gamma-ray Bursts (GRBs), rich and diverse X-ray afterglow components have been identified by a number of studies~\cite{burrows,nousek,obrien,willingale}. Often embedded within the X-ray lightcurves are X-ray flares (XRFs) in a large percentage of the GRBs~\cite{burrows,romano,falcone,chin}.\\ 

A number of studies suggest a connection between the prompt emission and the X-ray flaring activity. For example, the lag-luminosity relation for XRFs has been investigated by~\cite{Margutti} and was found to be consistent with the existing relation for the prompt emission~\cite{ukwatta,norris}. A very similar study~\cite{sultana} makes a connection between the prompt emission data and the late afterglow X-ray data and suggests that the lag-luminosity relation is valid over a time scale well beyond the early steep-declining phase of the X-ray light curve.~\cite{maxam} present a summary of the salient properties of XRFs and also show, using an internal shell collision model, that the main time histories of XRFs can be explained by the late activity of the central engine. Another study that hints at a connection between the prompt emission and the X-ray afterglow is that of~\cite{kocevski} in which the authors examined the evolution of pulse widths of the flares and found that the correlation between the widths of the pulses and time is consistent with the effects of internal shocks at ever increasing collision radii. In this work we focus on several temporal properties of the prompt emission and flaring emissions as seen especially in long bursts.

\section{Data And Methodology}

Following the work of~\cite{maclachlan13}, we have used a technique based on wavelets to extract a minimum time scale (MTS) for a sample of GRBs. 
The MTS determines the time scale at which scaling processes dominate over random noise processes. For the extraction of X-ray light curves, we used the
method developed by~\cite{evans}. Using the available software, we extracted X- ray-flare light curves with different time bins. By constructing log-scale 
diagrams (log (variance) of signal vs. inverse frequency in octaves) for the sample, we have determined the minimum time scale. An example of a log-scale 
diagram (in addition to the light curve) for an X-ray flare is shown in Figure 1. We also studied the extent to which the extraction of the MTS is sensitive to detector 
thresholds. Shown in Figure 2 is a result of a simulation of the extraction of the MTS for a number of bright GRBs. The figure shows MTS in octaves (inverse frequency) vs. Brightness (cast as signal-to- noise ratio, and defined as $\xi$ in~\cite{maclachlan13}). The input octaves are indicated as horizontal colored lines corresponding to the two selected octaves (6 and 7). The prompt emission sample lies in the brightness range of 0.3 $< \xi <$ 0.74, corresponding roughly to the range indicated by the horizontal double-headed black arrow. The extracted (octave) values, shown as blue squares and red circles, match well with the input values, indicating little dependence on brightness. Black triangles correspond to XRF data. The majority of the XRFs lie in a region of higher brightness (with a typical value of $\xi$ $\sim$ 3.0) compared to the prompt emission (with $\xi$ $\sim$ 0.5). This shows that the signal-to-noise ratio is significantly different for XRFs compared to that of the prompt emission, and in addition, further illustrates that the extracted MTS varies little with brightness.

\begin{figure}[htp]
\begin{minipage}[b]{0.5\linewidth}
\hspace{-1.6cm}
\centering
\fontsize{10}{12}\selectfont
\includegraphics[scale=0.31,angle = 0]{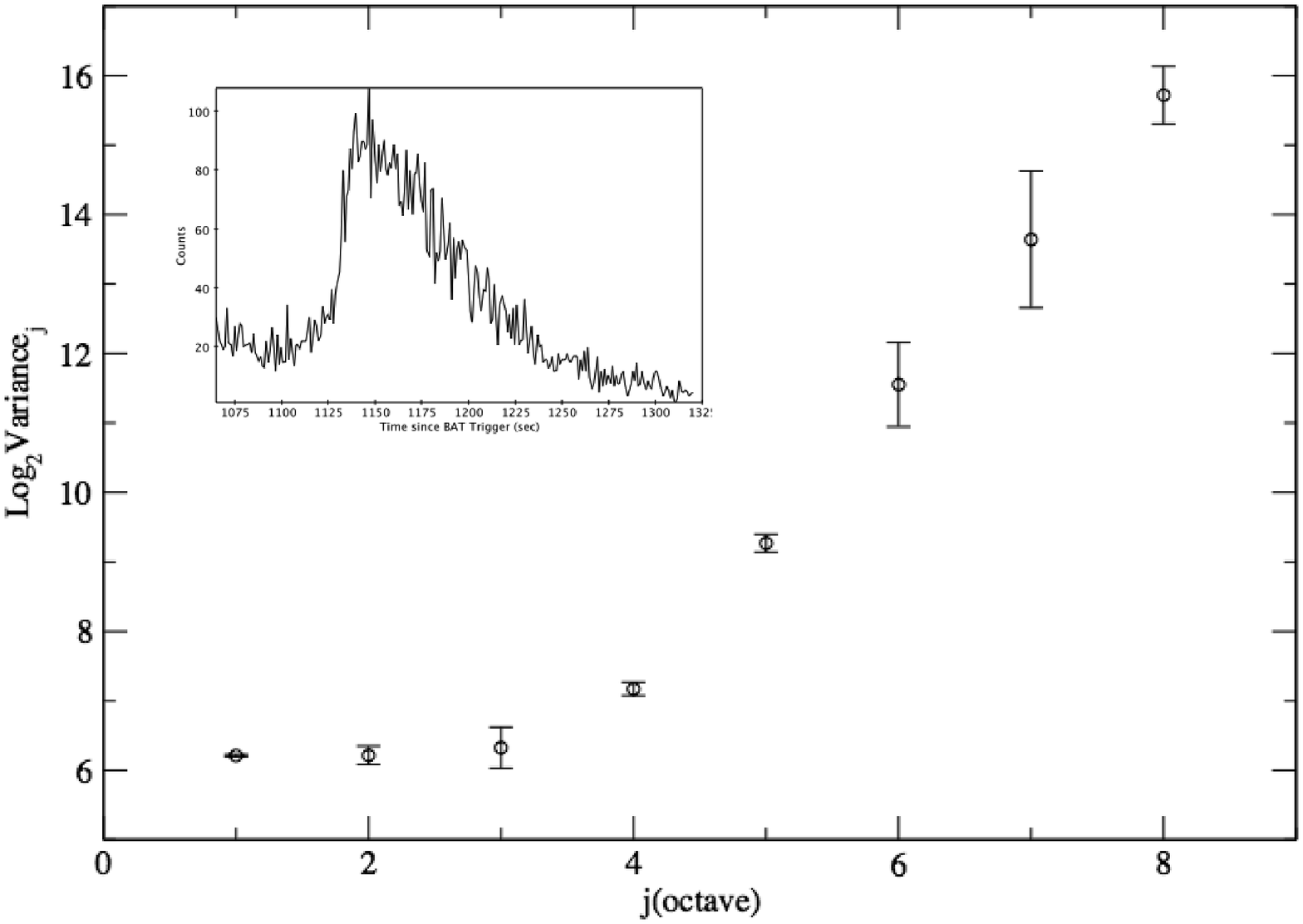}
\caption{Logscale diagram (and light curve) for the bright X-ray flare in GRB070520B: Log(Variance) of signal as a function of octave (inverse frequency). 
Plateau region is white noise and the sloped region is red noise.}
\label{fig1}
\end{minipage}
\hspace{0.12cm}
\begin{minipage}[b]{0.5\linewidth}
\centering
\includegraphics[scale=0.65,angle = 0]{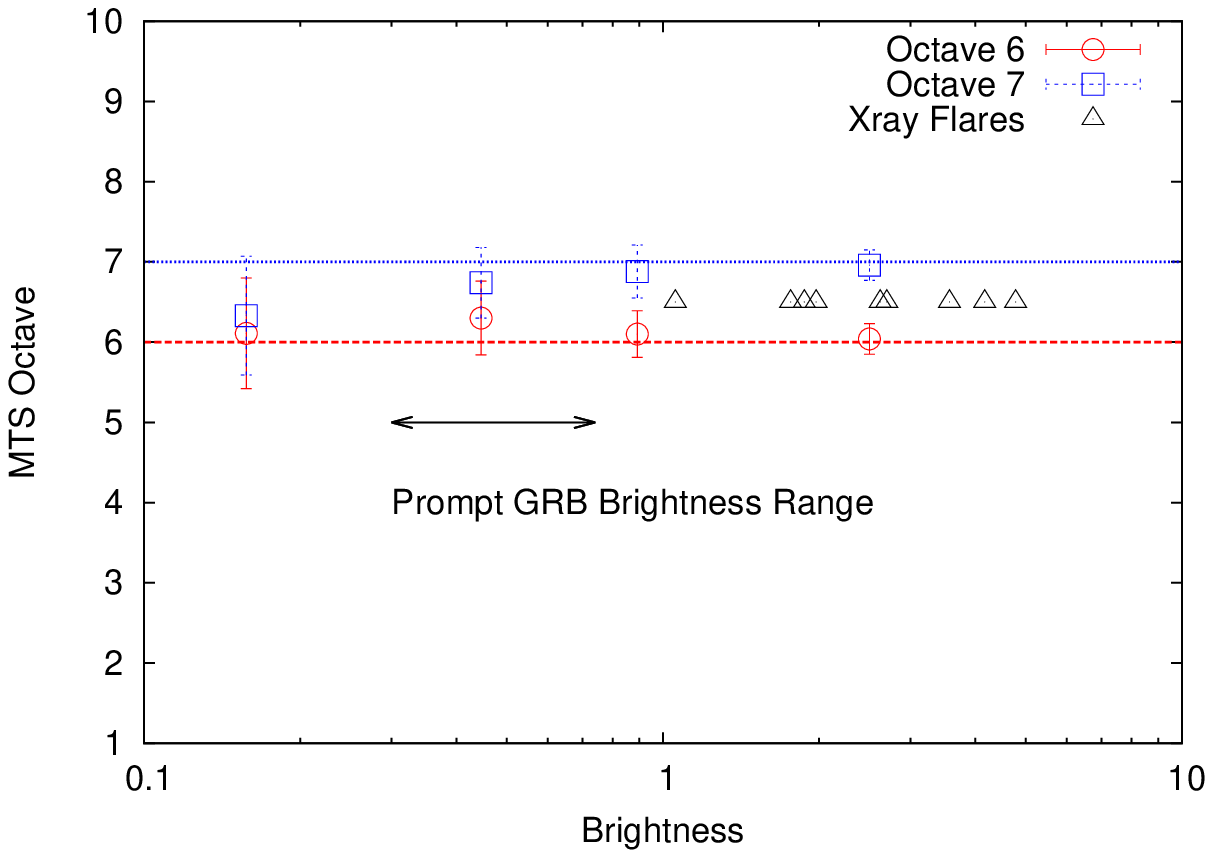}
\caption{MTS vs. Brightness for a sample of GRB prompt emission and XRFs. Input octaves shown as colored lines and extracted ones indicated by colored squares 
and circles. Triangles are XRFs.}
\vspace{0.5cm}
\label{fig2}
\end{minipage}
\end{figure}

\section{Results}

Using the extracted MTS and pulse-fit parameters (taken from literature), we show that a correlation exists between the MTS and the pulse-rise times. 
The correlation extends several decades of variability and includes the XRFs. This result is depicted in Figure 3, which shows pulse rise-times vs. MTS. 
Black data points indicate the prompt emission (with the pulse-fit parameters from~\cite{bhat}; the blue and green points depict the XRF data with pulse-fit 
parameters taken from~\cite{kocevski,Margutti} respectively. Also shown in the figure is a line depicting the equality of time scales. 
The best-fit line (not shown) leads to a slope of 1.26 $\pm$ 0.05. The Spearman correlation is 0.96 $\pm$ 0.02 and the Kendall correlation is 0.79 $\pm$ 0.02.

\begin{figure}[htp]
\begin{minipage}[b]{0.5\linewidth}
\hspace{-1cm}
\centering
\fontsize{10}{12}\selectfont
\includegraphics[scale=0.28,angle = 0]{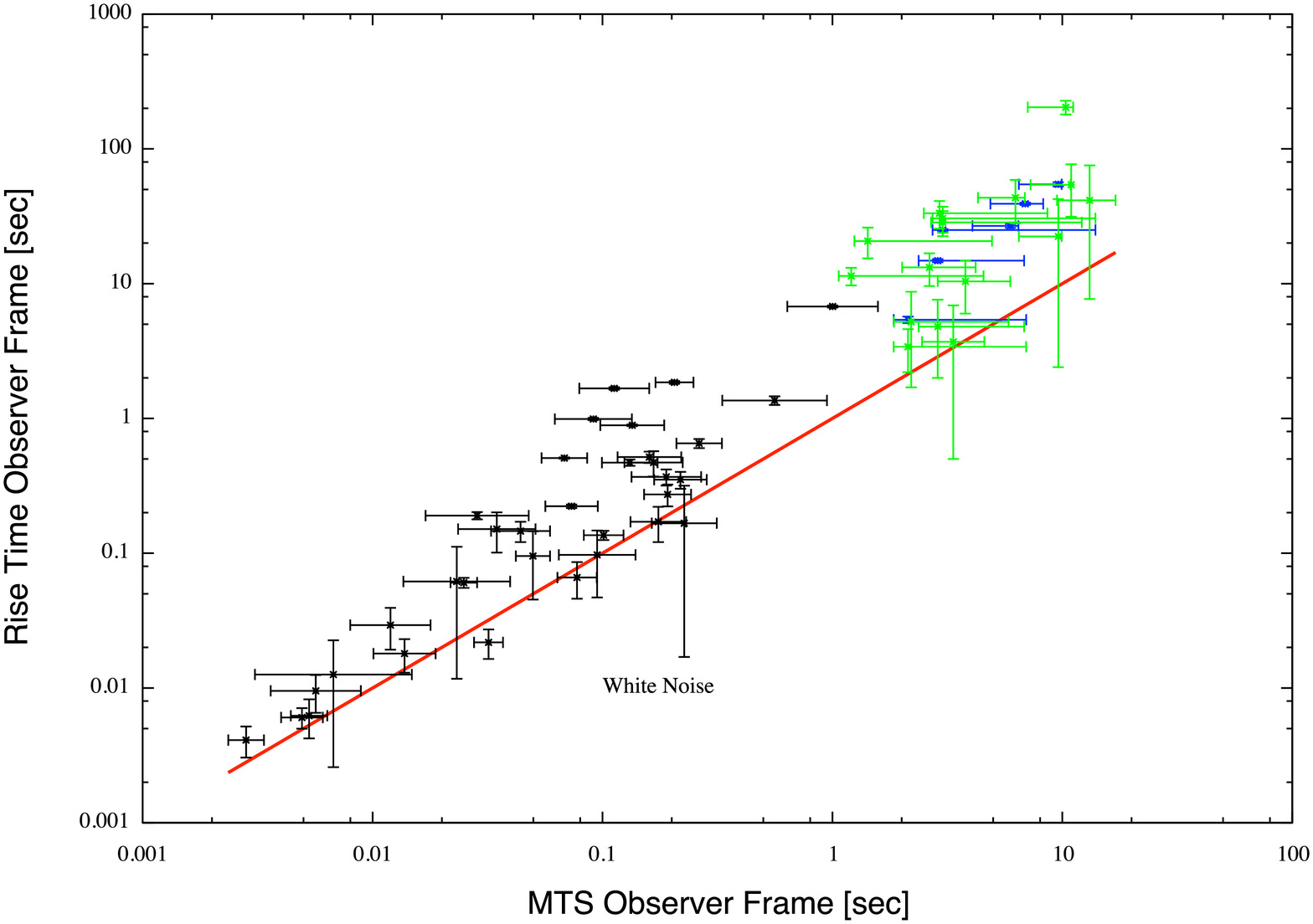}
\caption{Observer frame Pulse rise-times vs. MTS: Black points (prompt emission); green and blue points (XRF data). The red line indicates the equality of the respective 
temporal scales.}
\label{fig3}
\end{minipage}
\hspace{0.1cm}
\begin{minipage}[b]{0.5\linewidth}
\centering
\includegraphics[scale=0.27,angle = 0]{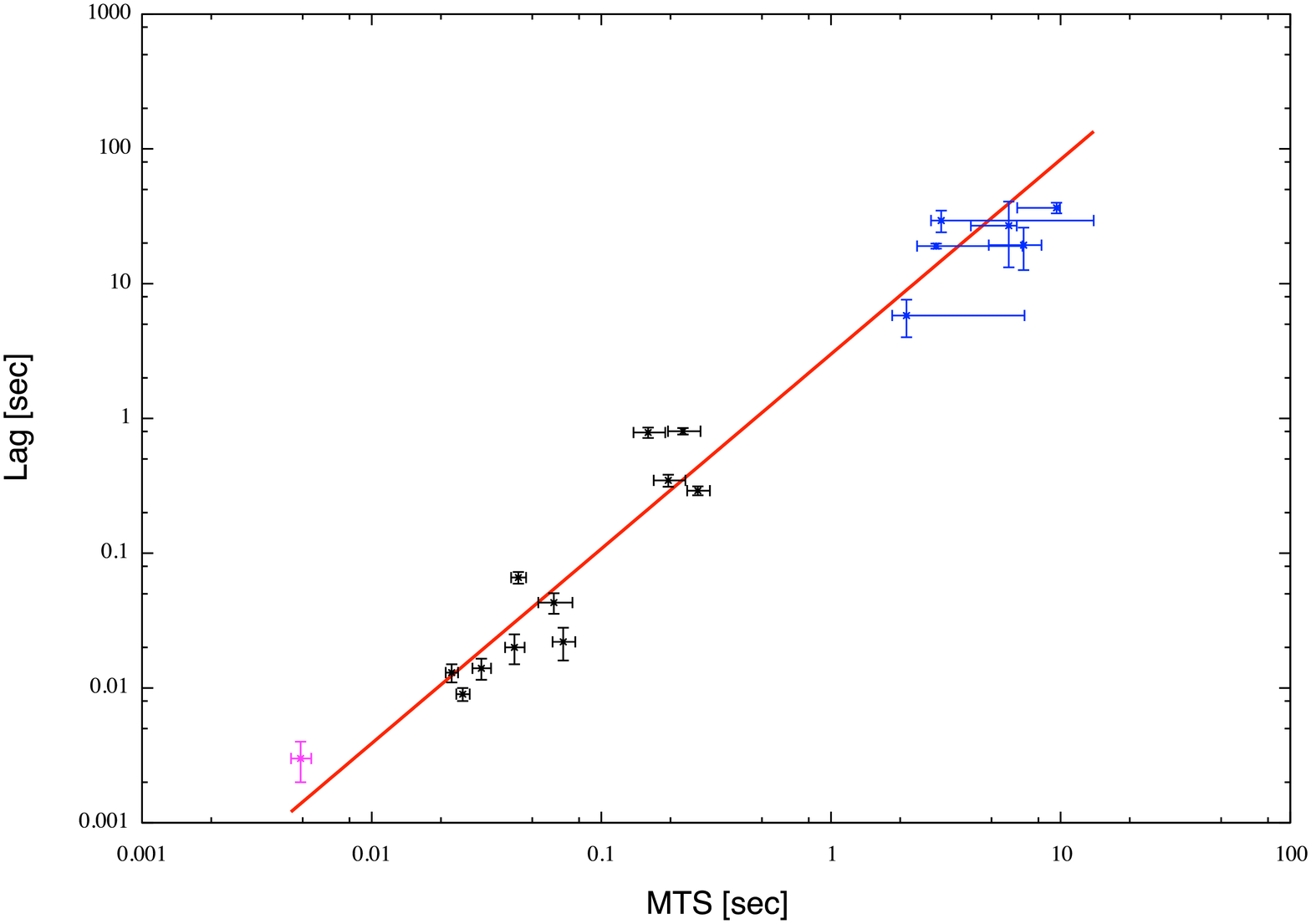}
\caption{Observer frame Spectral lags vs. MTS: Black points (prompt emission for long bursts); magenta point prompt emission for short burst); and blue points (XRF data). 
The red line indicates the best-fit to the data.}
\label{fig4}
\end{minipage}
\end{figure}

This result extends the work of~\cite{maclachlan12}, who examined prompt emission only, to the temporal domain covered by XRFs and reinforces their main conclusion that 
the two techniques, wavelets and pulse-fitting, can be used independently to extract a minimum time scale for physical processes of interest as long as close attention is 
paid to time binning and the proper identification of distinct pulses.\\ 

In order to pursue the apparent connection between the temporal properties of prompt emission and the XRFs further, we explore below the possible link between another temporal property, that of spectral lags, and the MTS. For the prompt emission data, we extracted spectral lags for various observer-frame energy bands using the CCF method described in detail by~\cite{ukwatta}. Some of these results have been presented by~\cite{sonbas}. Using the flare peak times reported by~\cite{Margutti}, we have also extracted the spectral lags for the XRFs between the energy bands 0.3-1 keV and the 3-10 keV respectively. A plot of the spectral lags vs. the MTS is shown in Figure 4. Black and magenta data points depict the prompt emission for long and short bursts; the blue points represent the XRF data. The red line indicates the best-fit (a slope of 1.44 $\pm$ 0.07) through the combined data set. The result clearly indicates a strong positive correlation (a Spearman correlation of 0.96 $\pm$ 0.05 and a Kendall correlation of 0.86 $\pm$ 0.05) between the two temporal features, spectral lag and the MTS.\\

The two correlations taken together i.e., the pulse-rise times vs. MTS and the spectral lag vs. MTS, are suggestive of more than a trivial connection between the prompt emission and the XRFs.

\section{Conclusions}
For a sample of long-duration GRBs detected by Fermi/GBM and Swift, we have extracted the minimum variability time scale (MTS) and spectral lags for both prompt emission and XRF light curves. We compare the MTS, extracted through a technique based on wavelets, both with the pulse rise times extracted through a fitting procedure, and spectral lags extracted via the CCF method. Our main results are summarized as follows;
\begin{itemize}
\item The prompt emission and the XRFs exhibit a significant positive correlation between pulse rise times and the MTS, with time scales ranging from several milliseconds to a few seconds respectively, and
\item The spectral lag for both the prompt emission and the XRFs shows a strong positive correlation with the MTS.
\end{itemize}
These results suggest a direct link between the mechanisms that lead to the production of XRFs and prompt emission in GRBs.

\end{document}